\documentstyle[multicol,pre,aps]{revtex}
\begin{document}
\input{epsf.tex}
\epsfverbosetrue

\title{Interaction of cavity solitons in
degenerate optical parametric oscillators.}
\author{Dmitry V. Skryabin
and William J. Firth}
\address{
Department of Physics and Applied Physics, University of
Strathclyde, Glasgow, G4 0NG, Scotland}

\date{\today}

\maketitle

\begin{abstract}
Numerical studies together with asymptotic and spectral analysis establish
regimes where soliton pairs in degenerate optical parametric oscillators fuse, 
repel, or form bound states. A novel bound state stabilized by coupled internal
oscillations is predicted.
\end{abstract}

\begin{multicols}{2}
\narrowtext

Many nonlinear media can support soliton-like
structures when contained in a driven optical cavity
\cite{rozanov,semi,staliunas,falkprl}.
We will refer to such
structures as {\em cavity solitons} (CS). In quadratic nonlinear media CS
have recently been predicted in both optical parametric oscillator
(OPO) \cite{staliunas} and second harmonic
generation \cite{falkprl} configurations. Although experimental observation of
$\chi^{(2)}$-CS remains a challenge, impressive bistability results
\cite{richy95josab} demonstrate the required level of nonlinearity and 
thus pave the way towards this goal.

The large values of effective
$\chi^{(2)}$ accessible in artificially phase-matched materials
in combination with their practically instantaneous 
response are  important advantages of using quadratic
nonlinearity for implementation of CS for all-optical
processing of information. They thus represent an
interesting alternative to the CS which can be created in cavities
with dispersive-absorptive \cite{rozanov},and resonant electron-hole
\cite{semi} types of nonlinearities. In all such schemes high CS density
is desirable and therefore understanding of their interaction is a practically
important  problem which is still largely open.
In this Letter we focus on the interactions of CS found in the below-threshold 
regime of a degenerate doubly resonant OPO,
under conditions where the signal
field has three coexistent plane-wave states\cite{staliunas}.

Assuming phase-matching, a plane-wave input field, and ignoring walk-off,
the mean-field OPO equations can be presented in the
following dimensionless form \cite{staliunas}
\begin{eqnarray}
&&-i\partial_tE_1=(\alpha_1\partial_x^2+\delta_1+i\gamma_1)E_1+
(E_2+\mu)E_1^*,\label{c1}\\
\nonumber
&&-i\partial_tE_2=(\alpha_2\partial_x^2+\delta_2+i\gamma_2)E_2+E_1^2/2,
\end{eqnarray}
 Here $E_1$ and $(E_2+\mu)$ are
the signal and pump fields, respectively, at frequencies $\omega$
and $2\omega$ (we use $\mu$ as a measure of the pump strength). The slow
time $t$ is scaled so that $\gamma_{m}$ (proportional to the
cavity damping rates) and $\delta_{m}$ (to the detunings from
its resonances) are of order unity. Here and below $m=1,2$.

This system can describe either diffractive or dispersive effects.
We consider $x$ a dimensionless transverse coordinate, and so 
set $\alpha_{m}=1/m$.  For this case, existence of CS for  $\delta_m<0$ was 
numerically demonstrated \cite{staliunas} for
$\mu_L<\mu<\mu_R$, where $\mu_L=|\gamma_1\delta_2+\gamma_2\delta_1|/
{\sqrt{\delta_2^2+\gamma_2^2}}$,
and $\mu_R={\sqrt{\delta_1^2+\gamma_1^2}}$ is the OPO threshold.
Within this range two different non-trivial homogeneous solutions
($E_m\ne 0$, $\partial_xE_m=0$) coexist with the trivial one ($E_m=0$), 
and the CS are sech-like localized states on the zero background.

We start our analysis by applying a perturbative method \cite{gorshkov}
to the problem of CS interaction. We seek solutions of Eqs. (1) in the form
\begin{eqnarray}
&&\nonumber
E_m(x,t)=A_m(x-x_A)+B_m(x-x_B)+\\
&& \epsilon (a_m(x-x_A,x_B,t)+b_m(x-x_B,x_A,t))+O(\epsilon^2),
\label{anzats1}\end{eqnarray}
where $A_m(x-x_A)$ and $B_m(x-x_B)$ are CS centred on $x_{A,B}$.  Note that 
Eqs. (\ref{c1}) are invariant with respect to a $\pi$ phase flip of the 
signal field, so that $A$ and $B$ can be either in-phase or out-of-phase CS.
We assume $0<\epsilon\ll 1$, and that the perturbation functions
 $a_m$, $b_m$ are negligible except close to $x_A$, $x_B$ respectively. 
We further assume that $x_{A,B}$ vary on the slow time scale $\tau=\epsilon t$
and that $d=|x_A-x_B|$ is large enough that the overlap functions
${\cal I}_1=A_2B_1^*+B_2A_1^*$ and ${\cal I}_2=A_1B_1$
are of order $\epsilon$. 

Substituting  ansatz (\ref{anzats1}) into Eqs.
(\ref{c1}) and truncating $O(\epsilon^2)$ terms we obtain two analogous systems of
equations for $a_m$ and $b_m$, the former expressible in the form:
\begin{equation}
(\hat {\cal L}_A-\partial_t)\vec a=-(\partial_{\tau}x_A)\vec\xi_A+\vec{\cal
I}/\epsilon,
\label{pert}\end{equation}
Here $\vec a=(Rea_{1},Rea_{2},Ima_{1},Ima_{2})^T$;
operator $\hat {\cal L}_A$ is the
linearization of Eqs. (\ref{c1}) around the soliton $A_m$;
$\vec \xi_A=\partial_x(ReA_{1},ReA_{2},ImA_{1},ImA_{2})^T$
is the neutral eigenmode of $\hat{\cal L}_A$ associated with translational
symmetry, $\hat {\cal L}_A\vec \xi_A=0$;
and $\vec{\cal I}=(-Im{\cal I}_1,-Im{\cal I}_2,Re{\cal I}_1,Re{\cal I}_2)^T$ 
controls the interaction of the two CS.

The solution of Eq. (\ref{pert}) should in general be expressed as a superposition
of the eigenmodes $\vec\xi_n$ of $\hat{\cal L}_A$, $\hat{\cal L}_A\vec\xi_n=\lambda_n\vec\xi_n$,
 with time dependent coefficients, because 
the CS interaction will couple to them all.
However, apart from the above-mentioned neutral eigenmodes, the only analytic knowledge
about the eigensystems of $\hat{\cal L}_A$ and $\hat{\cal L}_B$ is that they have
two bands of continuum modes with eigenvalues $\lambda$ lying on
$Re\lambda=-\gamma_m$, i.e. that all extended eigenmodes are damped.
We have obtained their full eigensystems numerically, using finite-differences, over wide
ranges of all relevant parameters.
We find that for sufficiently large dissipation all cavity solitons
are stable throughout the entire region of their
existence. A Hopf bifurcation can occur as photon lifetime is
increased, but we will not consider here any
parameter regions where isolated CS are unstable.
With oscillatory eigenmodes absent or well damped, only the
neutral mode is easily excited by external perturbations, and so we meantime 
neglect all other modes. This enables us to obtain semi-analytic results on CS interactions.

\begin{figure}
\setlength{\epsfxsize}{5.0cm}
\centerline{\epsfbox{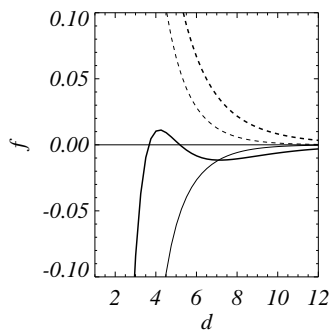}}
\caption{
Plots of the CS velocity function $f$ vs $d$. Full (dashed) lines correspond 
 to in-phase (out-of-phase) solitons and thin (thick) lines correspond to $\mu=1.6 (1.9)$.
Other parameters are  $\delta_2=-4$,  $\delta_1=-1.8$, $\gamma_1=1$, $\gamma_2=0.8$.
}
\label{friction}
\end{figure}

To exclude secularly growing solutions the right-hand side of Eq. (\ref{pert}) 
must be orthogonal to the neutral eigenmode of $\hat{\cal L}_A^{\dagger}$ 
(which we calculated numerically).  This solvability condition, together with that 
for the $B$ soliton, defines a function $f$ which governs the dynamical evolution 
of the distance $d$ between the soliton centers:
\begin{equation}
\partial_t d=f(d)\label{eq2}.
\end{equation}
We computed $f$ for both in-phase and out-of-phase interacting CS, for many 
parameter values.  Typical examples are plotted in Fig. \ref{friction}.
Regions where $f$ is negative (positive) correspond to CS
attraction (repulsion). Zeros of  $f(d)$ thus 
mark stationary {\em bound states} of CS pairs, which are stable if
$\partial_d f<0$ where $f=0$.

We find that this equation gives generally correct
predictions of the inter-soliton forces, in particular that in-phase CS 
attract and out-of-phase CS repel.  Both repulsion and attraction become
stronger as $\mu$ increases, presumably because the signal component
($E_1$) of the CS becomes less localized as $\mu$ approaches plane-wave 
threshold at $\mu_R$.  A similar effect can be envisaged in other CS models.
For in-phase CS the function $f$ can develop 
pairs of zeros, see Fig. \ref{friction}.  This predicts birth of new pairs of CS bound 
states, one stable and one unstable.

In Fig. 2 we present simulation results showing different
interaction scenarios for two CS initially separated by
about three soliton widths. First, we consider interaction of {\em in-phase} solitons. 
For small $\mu$ mutual attraction results in fusion of two
solitons into one (Fig. 2(a)). Gradually increasing $\mu$ we first
observe formation of a stable oscillatory bound state (Fig. 2(b)), then of a 
stationary bound state (Fig. 2(c)) which is stable (the radiation visible in Fig. 2(c)
 decays, albeit slowly).  Note that the equilibrium separation in Fig. 2(c) is predicted 
quite well by the appropriate zero of $f(d)$ in Fig. 1, even though these CS are close 
enough to endanger the assumptions of our perturbation method.
Stationary two-hump solitary states have been found previously 
\cite{staliunas} as solutions of an approximate equation derived from 
Eqs. (\ref{c1}), but no analysis of soliton interactions was performed.
Note that we have  found not only two-hump but also 
multi-hump solitary states. However the latter were usually dynamically unstable.
Further details on this issue  will be reported elsewhere.
Close to the upper boundary of CS existence the interaction of two solitons 
excites a global pattern, see Fig. 2(d), via generation of a switching wave from the 
stable trivial solution up into the modulationally unstable nontrivial homogeneous state.  
As predicted by Eq.  (\ref{eq2}), 
{\em out-of-phase} CS repel each other throughout the entire region 
of their existence - contrast Fig. 2(e) with Fig. 2(d), which corresponds to the 
same value of $\mu$.

Now we will describe numerical results of the interaction of CS where
weakly-damped oscillatory modes strongly influence the soliton interactions.
Oscillating solitons generally radiate energy, which can become trapped between 
neighbouring solitons, exerting a radiation force which may lead to formation of a 
bound state.  An effect of this kind has been reported for solitons in models with 
a weak global dissipation \cite{radiation}.
We investigated a quite different situation, where
linear waves escaping from the soliton are strongly damped.
Here strong interaction between the solitons is due, not to radiation modes,
but to proto-Hopf modes, and thus has novel aspects.

The effect is strong providing that two conditions are satisfied.  First, and crucially, 
the corresponding eigenmodes must have tails with well pronounced and weakly 
decaying oscillatory structure, see Fig. 3(a).  Second, as might be
expected, the oscillatory mode should be weakly damped (see Fig. 3(b)), i.e. the CS 
is close to a Hopf instability.  If both conditions hold, then, even if the global 
damping due to the $\gamma_m$ is strong, a CS acts as a guide 
for waves weakly damped in both space and time.  If a second CS is close enough, 
these guided waves can couple and reinforce each other.  Fig. 3(c, d) illustrates the
dynamics of two interacting CS having eigenmodes shown in Fig. 3(a).
Note that the separation of the interacting solitons in Fig. 3(c) is much
greater than their width.  Comparison with Fig. 3(b) clearly indicates that the
undamped pulsations shown in  Fig. 3(d) originate from coupling and mutual 
reinforcement of the oscillatory modes of the two solitons.
A further interesting point is that we find these dynamic 
bound states also for {\em out-of-phase} solitons, balancing the 
repulsion induced by their neutral mode interaction.

Quadratic nonlinearity is also known to support solitons in free propagation geometry
and in particularly interaction of these solitons has recently been studied both
experimentally  and theoretically \cite{george,sasha}. Hamiltonian nature of free
propagating solitons results in  their interaction obeying the laws of 
Newtonian dynamics \cite{sasha}.
Another important difference is that the relative phase of the interacting solitons can take only
two discrete values in a cavity while it is a continuous free parameter in a propagation scheme. 
In spite of these differences fusion of the in-phase solitons and repulsion of the out-phase
are common features in both schemes. However, existence of the stationary and oscillatory
bound states coupled either via translational neutral modes or via internal oscillatory modes
are novel important features arising due to presence of the  external pump and 
cavity losses. Balance between  the pump and losses acts as an additional, equally important with the
balance between diffraction and nonlinearity, mechanism of soliton formation inside an optical cavity.

In summary, we have presented the analytical and numerical study of the 
interaction of cavity solitons in a degenerate OPO and identified distinct static 
and dynamic binding mechanisms.

D.V.S. thanks  C. Etrich, F. Lederer, D. Michaelis and U. Peschel for warm
hospitality and illuminating discussions of many relevant questions during his
short visit to Jena. He also acknowledges financial support from the Royal
Society of Edinburgh and British Petroleum. The work is partially supported by
ESPRIT project PIANOS and EPSRC grant GR/M19727.


\begin{figure}
\setlength{\epsfxsize}{9.0cm}
\centerline{\epsfbox{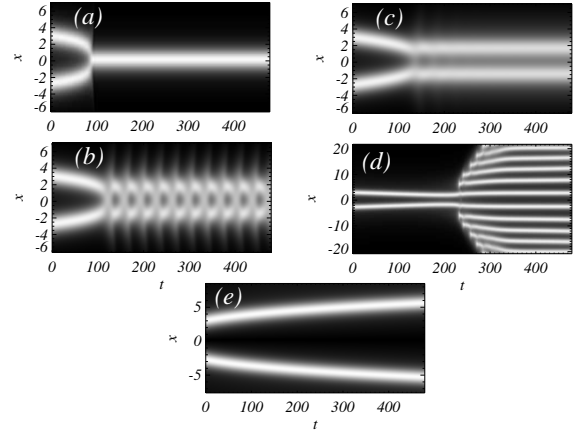}}
\caption{
Interaction dynamics of $\chi^{(2)}$ cavity solitons [9].
At different values of pump parameter $\mu$, in-phase CS: 
(a) merge, $\mu=1.6$; (b) form oscillatory bound state, $\mu=1.8$;
(c) form stable stationary bound state, $\mu=1.9$;
(d) generate a pattern {\em via} a switching wave, $\mu=2$.
Out-of-phase solitons repel, e.g. at $\mu=2$, (e).
Other parameters as for Fig. 1.}
\label{interaction}
\end{figure}

\begin{figure}
\setlength{\epsfxsize}{8.0cm}
\centerline{\epsfbox{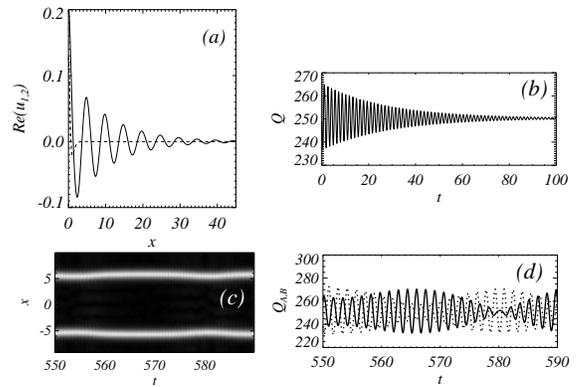}}
\caption{
Dynamic interaction of CS for $\mu=2$, $\delta_{1}=-3$,
$\delta_2=-12$, $\gamma_1=0.3$, $\gamma_2=1$, for which the CS has a mode with 
eigenvalue pair $\lambda\simeq -0.03\pm i4.14$: (a) Spatial structure of the eigenmode, 
$Re(u_1)$ - full lines, $Re(u_2)$ -dashed lines;
(b) Temporal evolution of signal energy $Q=\int dx|E_1|^2$
for slightly perturbed single soliton, showing damped oscillation;
(c) Spatio-temporal  evolution [9] of $|E_1|$ (time window much later than in (b)), 
showing dynamic bound state;
(d) Temporal evolution of signal energies of the two CS in (c), in the same time window, 
showing rapid undamped oscillations and slow energy exchange between the two solitons.
}\label{fig7}
\end{figure}
\end{multicols}

\begin{references}
\bibitem{com}
Only the central third of the computational
window in $x$ is shown in Figs. 2(a)-(c),(e), 3(c).
\end{references}
\end{document}